\documentclass[twocolumn,showpacs,preprintnumbers,amsmath,amssymb,
superscriptaddress,nofootinbib]{revtex4}
\usepackage{graphicx}
\newcommand{\eps}{\varepsilon}

\newcommand{\BM}[1]{{\mbox{\boldmath $#1$}}}

\newcommand{\be}{\begin{equation}}
\newcommand{\ee}{\end{equation}}
\newcommand{\ba}{\begin{eqnarray}}
\newcommand{\ea}{\end{eqnarray}}

\newcommand{\hh}{\, ,\hspace{0.5cm}}

\newcommand{\eq}[1]{(\ref{#1})}

\newcommand{\n}[1]{\label{#1}}

\begin{document}
\title{Stationary strings and branes in the higher-dimensional
Kerr-NUT-(A)dS spacetimes}

\author{David Kubiz\v n\'ak and Valeri P. Frolov}

\affiliation{Theoretical Physics Institute, University of Alberta, Edmonton,
Alberta, Canada T6G 2G7}

\email{kubiznak@phys.ualberta.ca}
\email{frolov@phys.ualberta.ca}

\date{November 14, 2007}  

\begin{abstract}
We demonstrate complete integrability of the Nambu-Goto equations
for a stationary string in the general  Kerr-NUT-(A)dS
spacetime describing the higher-dimensional rotating black hole.  The
stationary string in $D$ dimensions is generated by
a 1-parameter family of Killing trajectories and the problem of finding
a string configuration reduces to a problem of finding a
geodesic line in an effective $(D-1)$-dimensional space. Resulting integrability of 
this geodesic problem is connected with the existence of hidden symmetries 
which are inherited from the black hole background.
In a spacetime with $p$ mutually commuting Killing vectors it is possible to introduce a
concept of a $\xi$-brane, that is a $p$-brane with the worldvolume generated
by these fields and a 1-dimensional curve. We
discuss integrability of such $\xi$-branes in the Kerr-NUT-(A)dS spacetime.
\end{abstract}

\preprint{Alberta-Thy-16-07}

\maketitle

\section{Introduction} 
There are several reasons why the problem of interaction of 
strings and branes with black holes attracted interest recently.
Fundamental strings and branes are basic objects in string theory
\cite{Pol}, and black holes (as well as other black objects) form an
important class of solutions of the low-energy effective action in
this theory (see, e.g., \cite{Ort}). On the other hand, cosmic strings
and domain walls are topological defects which can be naturally
created during phase transitions in the early Universe (see, e.g.,
\cite{VS, Pol1, DK}). Their interaction with astrophysical black holes may
result in interesting observational effects. In both cases we are
dealing with a  problem when the interacting objects are
non-local and relativistic. An important example is an interaction of
a bulk black hole with a brane representing our world in the brane
world models (see, e.g., \cite{BR}). A stationary test brane interacting
with a bulk black hole can be used as a toy model for the study of 
(Euclidean) topology change transitions \cite{Toy}. This model
demonstrates interesting scaling and self-similarity properties
during such phase transitions, similar to the Choptuik critical
collapse  \cite{Chop} and merger black hole transitions \cite{Kol}. These
models may also have far going interesting consequences for the study of
phase transitions in quantum chromodynamics (see, e.g., \cite{CHD1,CHD2}).  

Even in an idealized case, when one neglects the effects connected
with the thickness of the strings and branes and their tension, this
problem is quite complicated. The reason is evident: the
Dirac-Nambu-Goto action for these objects in an external gravitational
field is very nonlinear. In a general case numerical calculations
are required (see, e.g., \cite{CSBH}). When the effects of
thickness and tension are taken into account these numerical
calculations become even more involved (see, e.g., \cite{Mori,Fl}).

Study of {\em stationary} configurations of strings and branes in a
background of a stationary black hole is simpler problem which
in several cases allows complete solution. One of the examples
is a stationary string in the Kerr spacetime. It was shown
\cite{FSZH} that the Hamilton-Jacobi equation for such a string
allows a complete separation of variables. It was also demonstrated 
\cite{CaFr} that this property is directly connected with the
hidden symmetry of the Kerr metric generated by the Killing tensor
\cite{WP} discovered by Carter in 1968 \cite{Car:68}. More recently,
Carters's method was
applied to 5-dimensional rotating black holes and the Killing tensor
was found in these spacetimes \cite{FS}. This result was used
to show that the equations for a stationary string in
the 5-dimensional Myers-Perry metric 
are completely integrable \cite{FrSte}.

In the present paper we demonstrate that this result allows a
generalization to  higher-dimensional rotating black holes
in an arbitrary number of spacetime dimensions.
Namely, we show that a stationary string configuration is
completely integrable in the general Kerr-NUT-(A)dS spacetimes
\cite{CLP}. We use the fact that after performing a dimensional reduction
along the Killing trajectories, the stationary string equation in a
$D$-dimensional stationary spacetime can be reduced to a geodesic
equation in a $(D-1)$-dimensional space with a metric conformal to the
reduced metric. The separability of the Hamilton-Jacobi equation in
this effective metric follows from the separability of the
Hamilton-Jacobi equation in the original $D$-dimensional
Kerr-NUT-(A)dS spacetime proved in \cite{FrKrKu} and a special
property of the primary (timelike) Killing vector. 

There is a natural generalization of the concept of a stationary
string in the case when there exist several mutually commuting
Killing vectors. If $p$ is a number of these fields  one  may consider
a $(p+1)$-hypersurface generated by the Killing vectors passing
through a 1-dimensional line. We call a $\xi$-brane an extended
object, a $p$-brane, with the worldvolume associated with this
hypersurface. We discuss integrability conditions for $\xi$-branes in
the Kerr-NUT-(A)dS spacetimes \cite{CLP} and give some examples of
integrable systems.

\section{Stationary strings}
Consider a string in a stationary $D$-dimensional spacetime $M^D$. Let
$x^a$ ($a=0,\ldots,D-1$) be coordinates
in it and
\be\n{1}
ds^2=g_{ab}dx^{a}dx^{b}
\ee
be its metric. We denote by $\xi^{a}$ the corresponding Killing vector
which is timelike at least in some domain of $M^D$. We call the string
{\em stationary} if $\xi^a$ is tangent to the 2-dimensional
worldsheet $\Sigma_{\xi}$ of the string in this domain. In other words, the surface
$\Sigma_{\xi}$ is generated by a 1-parameter family of the Killing
trajectories (the integral lines of $\xi^a$). 

A general formalism for studying a stationary spacetime based on its
foliation by Killing trajectories was developed by Geroch
\cite{Geroch}. In this approach, one considers a set $S$ of the
Killing trajectories as a quotient space and introduce the structure of the differential
Riemannian manifold on it. The projector $h_{ab}$ onto $S$ is related
to the metric $g_{ab}$ as follows
\be\n{2}
g_{ab}=h_{ab}+\xi_{a}\xi_b/\xi^2\, .
\ee
In this formalism, a stationary string is uniquely determined by a
curve in $S$. 

The equation for this curve follows from the Nambu-Goto
action 
\begin{equation}
I=-\mu\int d^2\zeta\left|\gamma\right|^{1/2}\,.
\end{equation}
Here $\mu$ is the string tension. As it enters the Nambu-Goto action 
as a common factor, its value is not important and one
can always put $\mu=1$. The string worldsheet can be parametrized by
$x^a=x^a(\zeta^A)$, where $\zeta^A$ 
are coordinates on $\Sigma_{\xi}$, ($A=0,1$). We
denote by $\gamma_{AB}$ the induced metric on $\Sigma_{\xi}$
\be
\gamma_{AB}={\partial x^a\over \partial{\zeta^A}}{\partial x^a\over
\partial{\zeta^B}}\,g_{ab}\, ,
\ee
and by $\gamma$ its determinant.

Let Killing time parameter be $t$, so that
$\xi^a\partial_a=\partial_t$, and let $y^i$ be coordinates which are
constant along the Killing trajectories (coordinates in $S$). Then, 
the non-vanishing components of the projection
operator $h_{ab}$ are $h_{ij}$ (reduced metric) and the metric \eq{1}-\eq{2} takes the form
\ba
ds^2&=&-F(dt+A_i dy^i)^2+h_{ij}dy^i dy^j\, ,\\
F&=&g_{tt}=-\xi_a\xi^a\hh A_i=g_{ti}/g_{tt}\, .
\ea
From \eq{2} it also follows that in these coordinates $h^{ij}=g^{ij}$.

We choose $\zeta^0=t$ and denote $\zeta^1=\sigma$. Then the string
configuration is determined by $y^i=y^i(\sigma)$. The induced metric is 
\ba
d\gamma^2=\gamma_{AB}d\zeta^Ad\zeta^b=-F(dt+A d\sigma)^2+dl^2\,,
\ea
where 
\be
dl^2=h d\sigma^2\,,\ A=A_i dy^i/d\sigma\,,\ 
h=h_{ij}{dy^i\over d\sigma}{dy^j\over d\sigma}\,, 
\ee
and it has the following determinant
\be
\gamma=\mbox{det}(\gamma_{AB})=-F h\, .
\ee
So, the Nambu-Goto action is 
\ba
I&=&-\Delta t E\,,\\
\label{E}
E&=&\!\mu\int\!\sqrt{F}dl=
\mu\int d\sigma
\sqrt{Fh_{ij}\frac{dy^i}{d\sigma}\frac{dy^j}{d\sigma}}\,\,.
\ea
In a static spacetime the equation \eq{E} has a very simple meaning: The energy density of a
string is proportional to its proper length $dl$ multiplied
by the red-shift factor $\sqrt{F}$.

The problem of a stationary string configuration therefore reduces to that of a geodesic in the  $(D-1)$-dimensional 
effective background
\begin{equation}\label{H}
dH^2=H_{ij}dy^i dy^j=F h_{ij} dy^i dy^j\,.
\end{equation}

In order to solve this geodesic problem we shall use the Hamilton-Jacobi method. That is, we 
shall attempt for the additive separation of the Hamilton-Jacobi equation 
\begin{equation}\label{HJ}
\frac{\partial S}{\partial\sigma}+H^{ij}\, \partial_i S\;\partial_j
S=0\, ,
\end{equation}
where $H^{ij}$ is the inverse of the effective metric \eqref{H} with 
the components given by
\begin{equation}\label{hij}
FH^{ij}=h^{ij}=g^{ij}.
\end{equation} 
If the Hamilton-Jacobi equation can be separated, the effective geodesic motion 
and hence also the stationary string configuration are completely integrable,
e.g., \cite{BeFr}.

\section{Stationary  strings in Kerr-NUT-AdS spacetime}
In this section we prove the complete integrability of a stationary string configuration
in the general Kerr-NUT-(A)dS spacetime \cite{CLP}.
After a suitable analytical continuation the metric takes the form\footnote{
The physical metric with proper signature is recovered when standard radial
coordinate $r=-ix_n$ and new parameter $M=(-i)^{1+\epsilon}b_n$ are introduced
(for more details see \cite{CLP}).
As these transformations do not affect the
discussed separability of the Hamilton-Jacobi equation we 
prefer to work with this more symmetric analytical continuation of the metric.  
}
\begin{equation}\label{metric}
\begin{split}
ds^2=&\,\sum_{\mu=1}^n\Bigl[\frac{dx_{\mu}^2}{Q_{\mu}}
  +Q_{\mu}\!\Bigl(\sum_{k=0}^{n-1} A_{\mu}^{(k)}d\psi_k\!\Bigr)^{\!2}\Bigr]\\
  &-\frac{\eps c}{A^{(n)}}\Bigl(\sum_{k=0}^n A^{(k)}d\psi_k\!\Bigr)^{\!2} \!,
\end{split}
\end{equation}
with $n=[D/2]$ and $\varepsilon=D-2n$.
Here,
\begin{gather}
A_{\mu}^{(k)}=\!\!\!\!\!\sum_{\substack{\nu_1<\dots<\nu_k\\\nu_i\ne\mu}}\!\!\!\!\!x^2_{\nu_1}\dots x^2_{\nu_k},\quad\!\!\! 
A^{(k)}=\!\!\!\!\!\!\sum_{\nu_1<\dots<\nu_k}\!\!\!\!\!x^2_{\nu_1}\dots x^2_{\nu_k}\;\label{co},\nonumber\\
Q_{\mu}=\frac{X_{\mu}}{U_{\mu}}\,,\quad
U_{\mu}=\prod_{\substack{\nu=1\\\nu\ne\mu}}^{n}(x_{\nu}^2-x_{\mu}^2)\,,\nonumber\\
X_{\mu}=\sum\limits_{k=\varepsilon}^{n}c_kx_{\mu}^{2k}-2b_{\mu}x_{\mu}^{1-\varepsilon}+\frac{\varepsilon c}{x_{\mu}^2}\,.
\end{gather}
Time is denoted by $\psi_0$, azimuthal coordinates by $\psi_k$,
${k=1,\dots,m=D-n-1}$, and ${x_\mu}$, ${\mu=1,\dots,n}$, stand for  
radial and latitude coordinates. 
Parameter $c_n$ is proportional to the cosmological constant \cite{HHOY}
\begin{equation}
{R_{ab}=(-1)^{n}(D-1)c_n\, g_{ab}}\,,
\end{equation}
and remaining constants $c_k$, $c$, and $b_{\mu}$ are related to rotation parameters, mass, and NUT 
parameters of the black hole (see \cite{CLP} for more details).
The inverse metric reads
\begin{equation}\label{invmetric}
\begin{split}
g^{ab}\partial_a\partial_b&=
\sum_{\mu=1}^n \frac1{X_\mu U_\mu}
\Bigl(\sum_{k=0}^m(-x_\mu^2)^{n-1-k}\partial_{\psi_k}\Bigr)^{\!2}\\
&+\sum_{\mu=1}^n Q_\mu(\partial_{x_\mu})^2
-\frac{\eps}{cA^{(n)}}\,(\partial_{\psi_n})^2\,.
\end{split}
\end{equation}

In the space with the metric \eq{metric} the vector
$\partial_{\psi_0}$, called {\em primary Killing}, plays a special
role. This vector (after the analytical continuation to the `physical'
spacetime) is timelike in the black hole exterior. It is also
directly connected with the principal Killing-Yano tensor of the
metric  \cite{KKPF}. We call a string stationary if it is tangent to
the primary Killing vector. For this string
one has
\ba\label{invH}
H^{ij}\partial_i\partial_j&=&\!F^{-1}\!\left[
\sum_{\mu=1}^n \frac1{X_\mu U_\mu}
\Bigl(\sum_{k=1}^m(-x_\mu^2)^{n-1-k}\partial_{\psi_k}\Bigr)^{\!2}
\right.\nonumber
\\
&+&\left. \sum_{\mu=1}^n Q_\mu(\partial_{x_\mu})^2
-\frac{\eps}{cA^{(n)}}\,(\partial_{\psi_n})^2
\right]\, ,\\
\label{F}
F&=&\sum_{\mu=1}^n Q_\mu -\frac{\eps c}{A^{(n)}}\,.
\ea
The expression in the square brackets of \eq{invH}, the reduced metric,
is similar to \eq{invmetric}. The only difference is that in the sum over $k$  
the term $k=0$ is omitted. This corresponds to the natural projection 
given by \eqref{hij}.

In the background of the metric $H_{ij}$ the Hamilton-Jacobi equation
\eqref{HJ} allows the additive  separation of variables 
\begin{equation}\label{sep}
S=w\sigma + \sum_{\mu=1}^n S_{\mu}(x_{\mu})+ \sum_{k=1}^{m} L_k\psi_k
\end{equation}
with functions ${S_\mu(x_\mu)}$ of a single argument ${x_\mu}$. 
Substituting \eqref{sep} into \eqref{HJ} we obtain
\begin{equation}\label{HJsep}
\begin{split}
Fw+\sum_{\mu=1}^{n}\frac1{X_\mu U_\mu }
\Bigl(\sum_{k=1}^m(-x_\mu^2)^{n-1-k}L_k\Bigr)^{\!2}\\
+\sum_{\mu=1}^{n} Q_{\mu}S_\mu'^2-\frac{\eps L_n^2}{cA^{(n)}}=0\,,
\end{split}
\end{equation}
where ${S_\mu}'^\;$ denotes the derivative of a function ${S_\mu}$
with respect to its single argument ${x_\mu}$. Using the explicit
form of $F$ and algebraic identity \cite{FrKrKu}:
\begin{equation}\label{Uids}
\frac1{A^{(n)}}=\sum_{\mu=1}^n \frac1{x_\mu^2U_\mu}\,,
\end{equation}
we can rewrite the last equation in the form
\begin{equation}\label{KGsep}
\sum_{\mu=1}^n\frac{G_{\mu}}{U_{\mu}}=0,
\end{equation}
where $G_{\mu}$ are functions of $x_{\mu}$ only:
\ba\label{Gmu}
G_{\mu}&=&X_{\mu}\left(S_{\mu}'^2+w\right)+\frac{1}{X_{\mu}}
\Bigl(\sum_{k=1}^m(-x_\mu^2)^{n\!-\!1\!-\!k}L_k\Bigr)^{\!2}\nonumber \\
&-&\frac{\eps \left(L_n^2/c+wc\right)}{x_{\mu}^2}\,.
\ea
Applying the Lemma proved in the Appendix of \cite{Krtous} we realize that the general solution 
of (\ref{KGsep}) is 
\begin{equation}
G_{\mu}=\sum_{k=1}^{n-1} C_{k}(-x_\mu^2)^{n\!-\!1\!-\!k}\;,
\end{equation}
where $C_k$ are arbitrary constants. 
So, we have obtained the equations for $S_{\mu}'$:
\ba\label{Scond}
S_\mu'^2&=&\ \frac{1}{X_\mu}\Bigl[\sum_{k=1}^{n-1} C_k\, (-x_\mu^2)^{n-1-k}+
\frac{\eps \left(L_n^2/c+wc\right)}{x_{\mu}^2}\Bigr]\nonumber
\\
&-&\frac1{X_\mu^2}\Bigl(\sum_{k=1}^m\bigl(-x_{\mu}^2\bigr)^{n-1-k}
L_k\Bigr)^{\!2}-w\, ,
\ea 
which can be solved by quadratures.

This completes the demonstration that in the general
higher-dimensional  rotating black hole spacetime \eqref{metric} the
reduced $(D-1)$-dimensional geodesic  problem \eqref{E} allows the
separation of the Hamilton-Jacobi equation \eqref{HJ} and therefore
the stationary string configuration  is completely integrable.

\section{Hidden symmetries}
The resulting complete integrability of the stationary string configuration
in the Kerr-NUT-(A)dS spacetime \eqref{metric} is connected with 
the existence of hidden symmetries of the 
$(D-1)$-dimensional effective metric $H_{ij}$.
Namely, there exist $(n-1)$ irreducible Killing tensors $C^{ij}_{(k)}$,
$(k=1, \dots, n-1)$, which 
give the constants of motion 
\begin{equation}
C_k=C_{(k)}^{ij}p_ip_j\,,
\quad D_{\!(m}C^{(k)}_{ij)}=0\,,
\end{equation}
and allow the separation of the Hamilton-Jacobi equation \eqref{HJ}
in the background $H_{ij}$. In the last formula $p_i=\partial_i S$
are the `momenta' of geodesic motion
and $D_i$ denotes the covariant derivative with respect to  
$H_{ij}$.

One can easily find the explicit form of $C^{ij}_{(k)}$ by
inverting \eqref{Gmu}. Let us multiply it by 
$A_{\mu}^{(l)}/U_\mu$, sum over $\mu$, and use identities
\cite{FrKrKu}:
\begin{equation}
\sum_{\mu=1}^{n} \frac{(-x_\mu^2)^{n\!-\!1\!-\!k}}{U_\mu}A_\mu^{(l)} 
= \delta_k^l\,,\quad
\sum_{\mu=1}^n \frac{A_{\mu}^{(k)}}{x_\mu^2 U_\mu}
=\frac{A^{(k)}}{A^{(n)}}\,,
\end{equation}
which are valid for $l,k=0,\dots,n-1$. Then we obtain
\ba
C^{ij}_{(k)}&=&K^{ij}_{(k)}-F_{(k)}H^{ij}\,,\\
F_{(k)}&=&\sum_{\mu=1}^n Q_\mu A_\mu^{(k)}-\frac{\eps c
A^{(k)}}{A^{(n)}}\, .
\ea
Here $K^{ij}_{(k)}$ are natural projections of the tensors
\begin{equation}\label{KT}
\begin{split}
K^{ab}_{(k)}\partial_a\partial_b
&=\sum_{\mu=1}^n
\frac{A_{\mu}^{(k)}}{Q_\mu U_\mu^2}\Bigl(\sum_{l=0}^m
(-x_\mu^2)^{n-1-l}\partial_{\psi_l}\Bigr)^{\!2}\\
&
+\sum_{\mu=1}^n A_{\mu}^{(k)}Q_\mu(\partial_{x_\mu})^2
-\frac{\eps A^{(k)}}{cA^{(n)}}(\partial_{\psi_n})^2\,.
\end{split}
\end{equation}
That is, similar to \eqref{invH},  the direction 
$\partial_{\psi_0}$ is projected out (the term $l=0$ is omitted).

In fact, the tensors $K^{ab}_{(k)}$, $(k=1,\dots,n-1)$, are 
the irreducible Killing tensors  for the 
Kerr-NUT-(A)dS metric \eq{metric} \cite{KKPF,FrKrKu}.
And so one can say that the hidden symmetries of the 
$(D-1)$-dimensional effective metric $H_{ij}$
are `inherited' from the hidden symmetries  of $g_{ab}$.

A nontrivial property which follows from the separability of the Hamilton-Jacobi equation  
(see, e.g., \cite{BeFr}) is that the constants $C_k$ mutually Poisson commute,
or equivalently, the Schouten brackets, in the background $H_{ij}$, of the corresponding Killing tensors vanish: 
 \begin{equation}
\bigl[{C}_{(k)},{C}_{(l)}\bigr]_{H}^{\,{ijm}}\!\!=
C_{(k)}^{n(i}D_n C_{(l)}^{jm)}-
C_{(l)}^{n(i}D_n C_{(k)}^{jm)}=0\,.
\end{equation}

Let us also mention that the projections $K^{ij}_{(k)}$
are the Killing tensors for the reduced metric $h_{ij}$ 
and obey
\begin{equation}
\bigl[{K}_{(k)},{K}_{(l)}\bigr]_{h}^{\,{ijm}}=0\,.
\end{equation}
These results can be easily obtained by separating 
the Hamilton-Jacobi equation in the background of the 
reduced metric $h_{ij}$. We expect them to be more general.
(For a discussion and necessary conditions regarding
the projection of a single Killing tensor see \cite{CaFr}.)

We have seen that the existence of the Killing tensors $C^{ij}_{(k)}$ for the metric 
$H_{ij}$ is the property inherited from the metric $g_{ab}$ \eqref{metric}.
This metric possesses even more 
fundamental symmetry---connected with the
principal Killing-Yano tensor \cite{KF} from which all the Killing tensors \eqref{KT}
are derivable \cite{KKPF}. A natural question arises whether
also $H_{ij}$ admits any (not necessary principal) Killing-Yano tensor.
 
In a general case the answer is negative. The necessary conditions for a   
Killing tensor in 4D to be the `square' of a Killing-Yano tensor
were  given by Collinson \cite{Col} (see also \cite{FeSa}). One can
easily check  that they are not satisfied and hence, at least in
$4D$, the metric $H_{ij}$ does not admit a Killing-Yano tensor.
In higher dimensions we can exclude the existence of the
`special' principal Killing-Yano tensor for the metric
$H_{ij}$.\footnote{The special principal
Killing-Yano tensor is a principal Killing-Yano tensor
obeying the additional properties as defined in \cite{HHHH1}.
It was demonstrated in \cite{HHHH2}
that the only higher-dimensional spacetime admitting this
special principal Killing-Yano tensor 
is the `generalized' Kerr-NUT-AdS spacetime, i.e. the 
spacetime different from $H_{ij}$.}

\section{$\xi$-branes}
In the above consideration we have focused on stationary strings, that is
strings generated by a 1-parameter family of timelike Killing
trajectories. There are two natural ways how one may try to generalize
this construction. First, one may consider other Killing vector
fields, and/or  second, in the case when there exist more than one Killing
vector, one may consider hypersurfaces formed by the set of Killing
trajectories passing through the same 1-dimensional curve.
Let us discuss these generalizations in more
detail.

For simplicity we assume that the spacetime $M^D$ allows $p$ mutually
commuting Killing  vectors which we denote by $\xi_{(M)}^a$,
($M,N=1,\ldots,p$). The Frobenius theorem implies that for each
point of the spacetime $M^D$ there exists (at least locally) a submanifold
of dimension $p$ generated by the Killing vectors $\xi_{(M)}^a$ passing
through this point. In other words, the set $\xi=\{\xi_{(M)}^a\}$ defines
a foliation of $M^D$. Similar to what was done in the Geroch
formalism for one Killing vector field, one can define a quotient
space $S$ of $M$ determined by the action of the isometry group
generated by $\xi$. This generalization of the Geroch's
formalism was developed in \cite{FW}. The metric $g_{ab}$ of the
spacetime $M^D$ can be written as
\ba
g_{ab}&=&h_{ab}+\Xi_{ab}\hh 
h_{ab}\xi^a_{(M)}=0\, ,\\
\Xi_{ab}&=&\sum_{M,N=1}^p a^{MN}\xi_{(M)a}\xi_{(N)b}\, .
\ea
Here $a^{MN}$ is the $(p\times p)$ matrix which is inverse to the
$(p\times p)$ matrix $a_{MN}=\xi_{(M)a}\xi^a_{(N)}$:
$a^{MN}a_{NK}=\delta^K_M$. A tensor $h_{ab}$ is a projection operator
onto $S$. 

Let us denote by $y^i$ $(D-p)$ coordinates which are constant along
the Killing surfaces generated by the set $\xi$, and by $\psi^M$ the
Killing parameters defined by the conditions
\be
\xi_M^a \partial_a=\partial_{\psi^M}\, .
\ee
The metric $g_{ab}$ in these coordinates $(x^a)=(y^i,\psi^M)$ takes the form
\be
ds^2=h_{i,j}dy^i dy^j+\!\!\sum_{M,N=1}^p\!\!
a^{MN}(\xi_{(M)a}dx^a)(\xi_{(N)b}dx^b)\, .
\ee
In these coordinates we also have
\be\label{NM}
a_{MN}=\xi_{(M)a}\xi^a_{(N)}=\xi_{(N)M}=\xi_{(M)N}\, .
\ee

A natural generalization of stationary strings $\Sigma_{\xi}$ are
$(p+1)$-dimensional objects  $\Sigma_{\xi}^p$ which are formed by a
1-parameter family of Killing surfaces.  We call them $\xi$-branes. 
In $(y^i,\psi^M)$-coordinates the equation of $\Sigma_{\xi}^p$ is
$y^i=y^i(\sigma)$. For this parametrization coordinates  on $\Sigma_{\xi}^p$ are 
$(\zeta^A)=(\psi^M,\sigma)$ ($A,B=1,\ldots,p+1$). The induced metric on
the $\xi$-brane takes the form
\ba\n{gamma}
d\gamma^2&=&\gamma_{AB}d\zeta^A d\zeta^B=(h+u)d\sigma^2\nonumber\\
&\hspace{-2.0cm}+&\hspace{-1.2cm}2d\sigma\!\sum_{M=1}^p \xi_{(M)\sigma}d\psi^M+
\!\!\!\sum_{M,N=1}^p a_{MN}d\psi^M d\psi^N.
\ea
Here we have defined
\be
\begin{split}
h=h_{ij}\frac{dy^i}{d\sigma}\frac{dy^j}{d\sigma}\,,\quad
\xi_{(M) \sigma}=\xi_{(M) i}\frac{dy^i}{d\sigma}\,,\\
u=\sum_{M,N=1}^p
a^{MN}\xi_{(M)\sigma}\xi_{(N)\sigma}\, .
\end{split}
\ee
In order to derive \eq{gamma} we used \eq{NM}. 

The metric $\gamma_{AB}$ can be considered as a block matrix of the
form
\be
{\BM \gamma}=\left(   
\begin{array}{cc}
A & B\\
C & D
\end{array}
\right)
\ee
where $A$ is a 1-dimensional matrix and $D$ is a matrix $(p\times p)$.
If $|Z|$ is a determinant of a matrix $Z$, then one has the following
relation for the determinant of a block matrix (see, e.g.,
\cite{Gant})
\be
\left|   
\begin{array}{cc}
A & B\\
C & D
\end{array}
\right|=|D||A-CD^{-1}B|\, .
\ee
Using this equation one obtains
\be
\gamma=\mbox{det} (\gamma_{AB})=\left|   
\begin{array}{cc}
h+u & \xi_{(M)\sigma}\\
\xi_{(N)\sigma} & a_{MN}
\end{array}
\right| = h\, {\cal F}_{\xi}\,,
\ee
where 
\begin{equation}
{\cal F}_\xi=\mbox{det}(a_{MN})=\mbox{det}(\xi_{(M)}^a\xi_{(N) a})
\end{equation}
is the Gram determinant for the set $\xi=\{\xi_{(M)}\}$ of the Killing vectors.

The Dirac-Nambu-Goto action for a $(p+1)$-dimensional brane is
\be
I=-\mu\int d^{p+1}\zeta \sqrt{|\gamma|}\, ,
\ee
where $\gamma$ is the determinant of the induced metric on the brane $\gamma_{AB}$.
For a $\xi$-brane this action reduces to the following expression\footnote{
In our derivation we have focused on a $1$-dimensional line in S generating $\xi$-branes.
The same construction remains valid for, let us say, 
$q$-dimensional hyperspace in 
S in the case of a $(p+q)$-dimensional brane. Then, denoting coordinates 
on the worldvolume of such brane by $(\zeta^A)=(\psi^M,\sigma^\alpha)$,
$(\alpha,\beta=1,\dots,q)$, and repeating the same steps one would obtain 
\be
\gamma={\rm det}(h_{\alpha\beta})F_\xi=hF_\xi,\quad 
h_{\alpha\beta}=h_{ij}\frac{dy^i}{d\sigma^{\alpha}}\frac{dy^j}{d\sigma^{\beta}}\,,
\ee
and
\ba
I&=&-\mu V {\cal E},\quad 
{\cal E}=\int\!\!\!\sqrt{{\cal F}_\xi} dv,\quad 
dv=\sqrt{h}d^q\!\sigma\,.
\ea
}

\ba
I&=&-\mu V {\cal E}\hh dl^2=h d\sigma^2\, ,\\
V&=&\int d^p \psi^N \hh
{\cal E}=\int \sqrt{{\cal F}_\xi} dl\, .
\ea
Thus after the dimensional reduction the problem of finding a
configuration of a $\xi$-brane  reduces to a problem of solving a
geodesic equation in the  reduced $(D-p)$-dimensional space
with the metric
\begin{equation}\n{HH}
dH^2=H_{\,ij}dy^idy^j={\cal F}_\xi h_{\,ij} dy^idy^j\, .
\end{equation}

If the original metric $g_{ab}$ admits a Killing tensor $K^{ab}$
then, since $h^{ij}=g^{ij}$, the natural projection $K^{ij}$ is
also a Killing tensor for the metric $h_{ij}$.  However, the full
effective metric $H_{ij}$ does not inherit this symmetry unless the
`red-shift' factor ${\cal F}_{\xi}$ is of the special `separable
form'.   Only then, the Hamilton-Jacobi equation \eqref{HJ} for the geodesic
motion in the metric \eq{HH} allows complete separation of
variables.

\section{$\xi$-branes in Kerr-NUT-AdS spacetime}

\subsection{Separability condition}

Let us discuss now the problem of integrability of $\xi$-branes in
the Kerr-NUT-(A)dS metric \eqref{metric}.  There we have $m+1$
Killing fields $\partial_{\psi_k}$, $k=0,\dots,m$ and we may choose
any arbitrary subset of them as the set $\xi$.  In general, however,
the corresponding red-shift factor ${\cal F}_\xi$ will not be of the
separable form.

More specifically, one requires that the red-shift factor can be written as 
\begin{equation}\label{rs}
{\cal F}_\xi=\sum_{\mu=1}^n \frac{f_\mu(x_\mu)}{U_\mu}\,,
\end{equation}    
with $f_\mu$ functions of $x_\mu$ only, 
in order to allow the separation of variables for the Hamilton-Jacobi equation in 
the effective background ${H}_{ij}$.
The corresponding Killing tensors $(k=1,\dots,n-1)$ would be then
\begin{equation}\label{notriv}
{C}_{(k)}^{ij}={K}_{(k)}^{ij}-f_{(k)}H^{ij}\,,
\end{equation}
where ${K}_{(k)}^{ij}$ are due natural projections of \eqref{KT}, with directions
from the set $\xi$ projected out, and
\begin{equation}
f_{(k)}=\sum_{\mu=1}^n\frac{f_\mu A_{\mu}^{(k)}}{U_\mu}\,.
\end{equation}

In the case of a stationary string, i.e. for $\xi=\{\partial_{\psi_0}\}$, 
the red-shift factor \eqref{F}, the norm of the 
primary Killing field $\partial_{\psi_0}$, possesses the property
\eqref{rs}, with
\begin{equation}
f_{\mu}=X_\mu-\frac{\epsilon c}{x_{\mu}^2}\,,
\end{equation}
and the integrability proved in the section III is justified.

\subsection{$\xi$-branes in 4D}
In 4D a stationary string is the only nontrivial example  of a
$\xi$-brane for which (in these coordinates) integrability can be
proved. Indeed, as discussed in \cite{CaFr} only in the
exceptionally  symmetric case of de Sitter space itself one can
obtain the integrability of the axially symmetric $\xi$-string with
$\xi=\{\partial_{\psi_1}\}$.\footnote {The asymmetry between the
Killing fields is connected with  the separability of the
Klein-Gordon equation, see, e.g., \cite{CaFr} and reference therein.
In higher-dimensional spacetime \eqref{metric} this separability was
demonstrated in \cite{FrKrKu}.    }

The last possibility of a $\xi$-brane in 4D 
Kerr-NUT-(A)dS spacetime is 
the axially symmetric stationary domain wall, 
$\xi=\{\partial_{\psi_0}, \partial_{\psi_1} \}$.
Let us consider this important example in more detail.
The action takes the form
\begin{equation}
I=-\mu \Delta \psi_0 \Delta \psi_1 
{\cal E}\,,\quad  
{\cal E}=\int\!d\sigma\sqrt{H_{ij}\frac{dy^i}{d\sigma}\frac{dy^j}{d\sigma}}\,,\quad 
\end{equation}
where the effective $2$-dimensional metric is
\begin{equation}
dH^2=H_{ij}{dy^i}{dy^j}=
{\cal F}_\xi\left(\frac{dx_1^2}{Q_1}+\frac{dx_2^2}{Q_2}\right).
\end{equation}
The red-shift factor reads
\begin{equation}
\begin{split}
{\cal F}_\xi=\left|   
\begin{array}{cc}
g_{\psi_0\psi_0} & g_{\psi_0\psi_1}\\
g_{\psi_0\psi_1} & g_{\psi_1\psi_1}
\end{array}
\right|=
\sum_{\mu=1}^2 \frac{f_{\mu}}{U_{\mu}}\,,
\end{split}
\end{equation}
where
\begin{equation}
f_\mu=x_\mu^2X_\mu(X_1+X_2).
\end{equation}
Evidently, $f_\mu$ becomes function of $x_\mu$ only in the case
when all parameters, but $c_0$, vanish. Only in that trivial case 
the Hamilton-Jacobi equation for the axially symmetric stationary 
domain wall in 4D can be separated.

The stationary string configuration remains the only one separable
also  in the standard  Boyer-Lindquist coordinates which can be
obtained from our coordinates  by the identifications given in
\cite{KuKr}.

\subsection{$\xi$-branes in 5D} 
In 5D the situation is more interesting.
There we can prove the integrability 
of the axisymmetric $\xi$-string, $\xi=\{\partial_{\psi_1}\}$, under the condition that $c_1=0$.
Indeed, then the red-shift factor takes the separable form \eq{rs} with
\begin{equation}
f_1(x_1)=2b_2x_1^4+cx_1^2\,,\quad f_2(x_2)=2b_1x_2^4+cx_2^2\,.
\end{equation}

Also, the axially symmetric stationary  
$\xi$-brane, $\xi=\{\partial_{\psi_0}, \partial_{\psi_1}\}$ is completely 
integrable in the case of a vacuum ($c_2=0$) 
5D spacetime \eqref{metric} with $c_1=0$.
In that case,
\begin{equation}
f_1(x_1)=4b_1b_2x_1^2+2cb_1\,,\quad f_2(x_2)=4b_1b_2x_2^2+2cb_2\,.
\end{equation}
In both cases the nontrivial Killing tensor responsible for the integrability 
is given by  \eqref{notriv}.

However restrictive and  unlikely
to be generally satisfied the condition \eqref{rs} seems,  the above examples
illustrate the special cases where
complete integrability of $\xi$-branes
can be analytically proved. We postpone the discussion of the
existence of other nontrivial examples elsewhere.

\section{Summary}
We have studied integrability of the Nambu-Goto
equations for a stationary string configuration near
a higher-dimensional rotating black hole.  In
a general stationary spacetime this problem reduces to finding a
geodesic in the effective  $(D-1)$-dimensional background ${H}_{ij}$.
In the Kerr-NUT-(A)dS spacetime \eqref{metric} the geodesic equation can be
integrated by separation of variables of  the corresponding
Hamilton-Jacobi equation. This separability  is a consequence of the
fact that ${H}_{ij}$ inherits some of the hidden symmetries of the
black hole.  Namely, it inherits $(n-1)$ irreducible mutually
commuting Killing tensors which  correspond to natural projections of
the Killing tensors present in ${g}_{ab}$. In a general
case there are no (antisymmetric)  Killing-Yano tensors generating
these (symmetric rank 2) Killing tensors.

The problem of integrating of equations for $\xi$-branes is more
complicated. We gave some examples where these equations are completely
integrable, but in the general case the complete integrability is not
possible. It would be interesting to find other, physically
interesting, examples of completely integrable $\xi$-branes in 
higher dimensional black hole spacetimes. It is also interesting to
study cases where there exist non-complete but non-trivial sets of
(quadratic in momenta) integrals of motion for $\xi$-branes related to
the hidden symmetries of the black hole background.

\section*{Acknowledgments}
One of the authors (D.K.) is grateful to the Golden Bell Jar Graduate
Scholarship in Physics at the University of Alberta and appreciates
the hospitality of the Department of Mathematics and Statistics at
the Dalhousie   University during the work on the paper. V.F. thanks
the Natural Sciences and Engineering Research Council of Canada and
the Killam Trust for the financial support.

\end{document}